\documentclass[prl,amsmath,superscriptaddress,citeautoscript,twocolumn,showpacs,floatfix]{revtex4}

\usepackage{amssymb}
\usepackage{graphicx}

\usepackage{mathrsfs}
\usepackage{amssymb,amsmath}
\usepackage{graphics}
\usepackage{bm}

\usepackage{epsfig}
\usepackage{color}
\usepackage{amsfonts}

\usepackage{amscd}

\begin{document}

\title{Intrinsic coherence dynamics and phase localization in Aharonov-Bohm Interferometers}

\author{Matisse Wei-Yuan Tu}
\affiliation{Department of Physics, National Cheng Kung University,
Tainan 70101, Taiwan}

\author{ Wei-Min Zhang}
\email{wzhang@mail.ncku.edu.tw} \affiliation{Department of Physics,
National Cheng Kung University, Tainan 70101, Taiwan}

\author{Jinshuang Jin}
\affiliation{Department of Physics, National Cheng Kung University,
Tainan 70101, Taiwan} \affiliation{Department of Physics, Hangzhou
Normal University,  Hangzhou 310036, China }

\begin{abstract}
The nonequilibrium real-time dynamics of electron coherence is
explored in the quantum transport through the double-dot
Aharonov-Bohm interferometers. We solve the exact master equation to
find the exact quantum state of the device, from which the changes
of the electron coherence through the magnetic flux in the
nonequilibrium transport processes is obtained explicitly.  We find
that the relative phase between the two charge states of the double
dot localizes to $\frac{\pi}{2}$ or $-\frac{\pi}{2}$ for all
different magnetic flux. This nontrivial phase localization process
can be manifested in the measurable occupation numbers.
\end{abstract}

\maketitle

\emph{Introduction.---} Quantum coherence and quantum transport in
mesoscopic electronic systems has attracted much attention due to
recent achievements in quantum technology. Controlling quantum
coherence in various quantum devices is essential to their
functioning.  Especially in a ring structured Aharonov-Bohm (AB)
interferometer where electron interference can be tuned by an
externally applied magnetic flux, a great amount of efforts has been
made to investigate its coherence transport properties, mostly in
the steady limit
\cite{YACOBY,JOE,HARRIS,LEVINSON,ALEINER,ADY,Hackenbroich01463,Koenig245301,Kon013855,Sigrist06036804,Entin02166801,Puller10256801}.
Alternatively we address here the real-time dynamics of building
electron coherence and then the subsequent evolution of the
intrinsic coherence in a two-terminal AB interferometer with two
quantum dots (QDs).

Coherence of electron transport through an AB interferometer has
long been characterized via conductance oscillation in magnetic
fields \cite{YACOBY,JOE,HARRIS}. Factors influencing AB oscillations
include "which path detection" \cite{YACOBY,LEVINSON,ALEINER},
electron-electron interaction \cite{ADY,Hackenbroich01463,Kon013855}
and inelastic scattering with phonons \cite{Sigrist06036804}, etc.
Extracting the transmission phase from AB oscillations has always
been the focus (see \cite{Entin02166801,Puller10256801} and
references there in) since it may contain information about
coherence of electron transport other than the interference by the
AB phase. For a double dot AB interferometer, what most intuitively
depicts the electron coherence other than AB interference is the
relative phase between the two dot charge states. This intrinsic
phase is generally entangled with the AB phase and its fundamental
dynamics has not been well understood so far. However, the dynamics
of this intrinsic phase is important for manipulating quantum
coherence in the application of quantum information processing.
Therefore in this Letter we shall directly study the nonequilibrium
electron dynamics to examine the effects of the magnetic flux on the
relative phase.

To explore the intrinsic coherence dynamics, we solve exactly the
time evolution of the electron charge states in the double dot AB
interferometer. We found that although different values of magnetic
flux will induce different relative phases initially, they will
eventually be localized to $\pi/2$ or $-\pi/2$. Remarkably, this
phase localization dynamics is reflected in the occupation number
rather than the transport current. For the latter, its magnetic flux
dependence has been extensively employed in characterizing the
coherence of electron transport.


\emph{Reduced density matrix and exact master equation.---} The
system under consideration is described by the Hamiltonian
consisting of three parts, ${\cal H}={\cal H}_{\rm s}+{\cal H}_{\rm
lead}+{\cal H}_{\rm c}$.  ${\cal H}_{\rm
s}=\sum_{ij}E_{ij}a^{\dag}_{i}a^{}_{j}$ where $i,j=1,2$ is the
Hamiltonian of the interferometer, with $a^{}_{i}$
($a^{\dagger}_{i}$) destroying (creating) an electron on dot $i$.
The lead Hamiltonian is ${\cal H}^{}_{\rm lead}=\sum_{\alpha \bm{k}}
\epsilon^{}_{\alpha \bm{k}}c^{\dag}_{\alpha \bm{k}} c^{}_{\alpha
\bm{k}}$, where $\alpha =L$ or $R$ and $c^{}_{\alpha \bm{k}}, \
c^{\dag}_{\alpha \bm{k}}$ the electronic operators of the leads, and
\begin{align}
{\cal H}_{\rm c}=\sum_{j \bm{k}}[V_{jL\bm{k}}
e^{i\phi_{jL}}a_{j}^{\dag}c^{}_{L \bm{k}}+V_{jR\bm{k}}
e^{i\phi_{jR}}c^\dag_{R \bm{k}}a^{}_j+{\rm H.c.}] \label{HTUN}
\end{align}
describes the coupling between the dots and the leads. In Eq.
(\ref{HTUN}), $\phi_{j\alpha}$ are the AB phase factors due to the
AB flux $\Phi$ such that
$\phi_{1L}-\phi_{2L}+\phi_{1R}-\phi_{2R}=\phi\equiv
2\pi\Phi/\Phi_{0}$ with $\Phi_{0}$ being the flux quantum. To single
out the effects of the threading magnetic flux, we will not include
explicitly the electron-electron interaction here.

Coherence between the two electron charge states of the double dot
is embedded in the reduced density matrix $\rho$ of the double dot
system, which can be obtained by tracing over the lead electrons
from the total reduced density matrix $\rho_{\rm tot}(t)=e^{-i{\cal
H}(t-t_0)} \rho_{\rm tot}(t_0) e^{i{\cal H}(t-t_0)}$.   The (exact)
equation of motion for $\rho$ has been derived in Ref.
\cite{Tu08235311,Jin10083013}. Assuming that initially, at
$t=t_{0}$, the dots are uncorrelated with the leads \cite{Leg871},
then
\begin{align}
&\dot{\rho}=-i[{\cal H}'_S(t),\rho(t)] \nonumber \\
& + \sum_{ij}\Big\{ \gamma_{ij}^{}(t)(2a^{}_{j}\rho(t) a^{\dag}_{i}-
a^{\dag}_{i}a^{}_{j}\rho(t)-\rho(t) a^{\dag}_{i}a^{}_{j})
\nonumber\\& + \widetilde{\gamma}^{}_{ij}(t)(a_{j}\rho(t)
a^{\dag}_{i} - a^{\dag}_{i}\rho(t) a^{}_{j}-
a^{\dag}_{i}a^{}_{j}\rho(t) +\rho(t) a^{}_{j}a^{\dag}_{i}) \Big\}\ .
\label{emaster}
\end{align}
Here, ${\cal H}'_S(t)=\sum_{ij} \epsilon'_{ij}(t)a^\dag_i a^{}_j$ is
the renormalized Hamiltonian of the double dot; ${\bm \epsilon}'$
and all other time-dependent coefficients in Eq.~(\ref{emaster}) are
given by
$\epsilon'^{}_{ij}(t) -i\gamma^{}_{ij}(t) = i[\dot{\bm u}\bm
u^{-1}]^{}_{ij}$,
$\widetilde{\gamma}^{}_{ij}(t)=[\dot{\bm u}\bm u^{-1}\bm v + {\rm
H.c.}-\dot{\bm v}]^{}_{ij}$ ,
where the matrix functions ${\bm u}$ and ${\bm v}$ obey the
following equations \cite{COM1}
\begin{subequations}
\label{uvn}
\begin{align}
\dot{\bm u}(\tau )+  i\bm E {\bm u}(\tau) & + \int_{t_0}^{\tau }
d\tau' \bm g (\tau-\tau') {\bm
u}(\tau')=0\ ,  \label{ue}\\
\dot{\bm  v}(\tau)+  i\bm E \bm  v(\tau) & + \int_{t_0}^{\tau }
d\tau' \bm g (\tau-\tau') \bm v (\tau') \notag \\ & =
 \int_{t_0}^{t }d\tau'
\widetilde{\bm g} (\tau-\tau')\bm u^\dag(t-\tau'+t_0)\ , \label{ve}
\end{align}
\end{subequations}
with the initial conditions $\bm u(t_{0})=I, \bm v(t_0)=0$. In
Eq.~(\ref{uvn}), $\bm E$ is the $2 \times 2$ energy matrix of the
double-dot, ${\bm g}$ and $\widetilde{\bm g}$ are the temporal
correlation functions resulted from the couplings to the leads via
the spectral density,
\begin{align}
\Gamma^{}_{\alpha  ij}(\omega )=2\pi\sum_{\bm {k}
\in\alpha}V_{i\alpha \bm{k}}V_{j\alpha
\bm{k}}e^{i(\phi_{i\alpha}-\phi_{j\alpha})}
\delta(\omega-\epsilon_{\alpha \bm k})\ ,\label{GAM}
\end{align}
such that \cite{Tu08235311}:
${\bm g}(\tau )=\sum_{\alpha =L,R}\int\frac{d\omega}{2\pi}
e^{-i\omega\tau}{\bm \Gamma}^{}_{\alpha}(\omega) $ and
$\widetilde{\bm g}(\tau )=\sum_{\alpha
=L,R}\int\frac{d\omega}{2\pi}f^{}_{\alpha}(\omega
)e^{-i\omega\tau}{\bm \Gamma}^{}_{\alpha}(\omega) $,
where $f_{\alpha}(\omega )=(\exp
[(\omega-\mu_{\alpha})/k_{B}T]+1)^{-1}$ is the initial Fermi
distribution for the electron reservoir (lead) $\alpha=L,R$.

\emph{Exact solution of the master equation.---} To monitor the
formation of coherence in the transport process, we prepare the
double dot with the empty state $|0\rangle$ at $t=t_{0}$.  We denote
the singly occupied states by $|1\rangle$ and $|2\rangle$ referring
to occupation of the first dot and the second dot, respectively, as
the two charge state basis. And the double occupancy is denoted by
$|3\rangle$. From the master equation (\ref{emaster}) with the above
preparation, the reduced density matrix becomes
\begin{align}
\rho(t)=\begin{pmatrix} \rho^{}_{00}(t) & 0 & 0 & 0 \\ 0 &
v_{11}(t)-\rho_{33}(t) & v_{12}(t) & 0 \\ 0 & v_{21}(t) &
v_{22}(t)-\rho_{33}(t) & 0 \\ 0 & 0 & 0 & \rho^{}_{33}(t)
\end{pmatrix} \ , \label{srdm}
\end{align} where the dynamics of the relative phase between the two charge states
is given explicitly by the off-diagonal matrix element
$\rho_{12}(t)=v_{12}(t)$, while $\rho_{00}(t)=\det[I-\bm v(t)]$ and
$\rho_{33}(t)=\det \bm v(t)$ determine the leakage effect and the
double occupation, respectively, and
$\rho_{ii}(t)=v_{ii}(t)-\rho_{33}(t)$ ($i=1,2$) is the probability
of each singly occupied charge state. It is obvious that the total
probability is conserved: ${\rm tr}\rho(t)=1$. The full reduced
density matrix is thus determined from $\bm v(t)$ which has a
general solution from Eq.~(\ref{ve}),
\begin{align}
\bm v(t)= \int \frac{d\omega}{2\pi}{\bm u} (t,\omega)\sum_\alpha
f_\alpha(\omega)\bm \Gamma_\alpha(\omega)  {\bm u}^\dag(t,\omega)\ ,
\label{svp}
\end{align} where ${\bm u}(t,\omega)\equiv \int_{t_0}^t
d\tau e^{i\omega(t-\tau)}\bm u(t-\tau+t_0)\ $.  It is then
sufficient to solve the first line of Eq.~(\ref{uvn}) for a full
understanding of the intrinsic coherence dynamics.

As being widely studied in the literature, with a degenerate double
dot given by the Hamiltonian ${\cal H}_{\rm
s}=\sum_{i}Ea^{\dag}_{i}a^{}_{i}$ and an equal coupling
$\Gamma_{L}=\Gamma_{R}=\Gamma/2$ in the white band limit, we can
well suppress the relaxation dynamics and exclusively focus on the
relative phase dynamics. Also, as a convention, we choose the gauge
$\phi_{1L}-\phi_{2L}=\phi_{1R}-\phi_{2R}=\phi/2$. These
considerations lead  to
$ \bm \Gamma_{L,R}={\Gamma_{}\over2} \begin{pmatrix} 1 & e^{\pm i \phi/2}\\
e^{\mp i\phi/2} & 1 \end{pmatrix}\ $. One can easily obtain from
Eq.~(\ref{ue}), by taking $t_{0}=0$,
\begin{align}
{\bm u}(t)=e^{-(iE+\frac{\Gamma}{2})t}\Bigl ({\rm
cosh}\frac{\Gamma_\phi t}{2} -\sigma^{}_{x}S(\phi){\rm
sinh}\frac{\Gamma_\phi t}{2}\Bigr ) \label{ug} ,
\end{align} where $\Gamma_{\phi}=\Gamma|\cos(\phi/2)|$ and $S(\phi)=\frac{\cos(\phi/2)}{|\cos(\phi/2)|}$.
Straightforwardly one can also find from Eq.~(\ref{svp})
\begin{align} {\bm v}(t)= v_0(t)I +
v_x(t)\sigma_x + v_y(t)\sigma_y + v_z(t)\sigma_z  ,\ \label{sov}
\end{align} where $v_0(t)=A^{}_{+}(t)+A^{}_{-}(t)$ and $\{v_x(t),
v_y(t), v_z(t)\}=\{S(\phi)(A^{}_{+}(t)-A^{}_{-}(t)),-{\rm
Re}B(t),S(\phi){\rm Im}B(t)\}$ with
\begin{align} & A^{}_{\pm}(t)=\frac{\Gamma}{4}(1 \pm
|\cos\frac{\phi}{2}|) \int\frac{d\omega}{2\pi}[f^{}_L(\omega) +
f^{}_R(\omega)]|u^{}_\pm(t,\omega)|^2
, \notag \\
& B(t)=\frac{\Gamma}{2}\sin\frac{\phi}{2}\int \frac{d\omega}{2\pi}
[f^{}_L(\omega)- f^{}_R(\omega)]u^{}_+(t,\omega)u^{\ast}_-(t,\omega)
,\label{ABcoef}
\end{align}
$u_{\pm}(t,\omega)=\frac{e^{i(\omega-E)t-\frac{1}{2}(\Gamma\pm\Gamma_{\phi})t}-1}{
i(\omega-E)-\frac{1}{2}(\Gamma\pm\Gamma_{\phi})}$.  The notation
$\sigma_{i}$ with $i=x,y,z$ denotes the Pauli matrices.

\emph{Phase localization.---} The electron coherence dynamics is
embedded in the two charge state density matrix [the central block
matrix in Eq.~(\ref{srdm})] which can be rewritten as
\begin{align} \label{qdm}
\rho_{\rm q}(t)=\frac{1}{2}\big[I + {\bm r}(t)\cdot {\bm
\sigma}\big]- \frac{\rho_{00}(t)+\rho_{33}(t)}{2}I \ ,
\end{align}
where ${\bm r}(t)=2\{v_x(t), v_y(t), v_z(t)\}$ is the polarization
vector of the two charge states, and $\rho_{00}(t)+\rho_{33}(t)$
clearly shows the leakage effect.  To exclusively study the
coherence between the two charge states, we shall apply a bias,
$\mu_{L}-E=E-\mu_{R}$ such that the two dots would be equally
occupied, $\rho_{11}(t)=\rho_{22}(t)$, i.e. $r_{z}(t)=0$ because
${\rm Im}(u^{}_+(t,\omega)u^{\ast}_-(t,\omega))$ is antisymmetric in
$\omega$ with respect to $E$.  The polarization vector then is fully
specified by $r_x(t)=2{\rm Re}\rho_{12}(t)=2v_x(t)$ and
$r_y(t)=-2{\rm Im}\rho_{12}=2v_y(t)$, which shows purely the
dynamics of coherence between the two charge states through the
relative phase $\varphi (t)$, defined explicitly by \begin{align}
\rho_{12}(t)=|\rho_{12}(t)|e^{i\varphi(t)}=\frac{1}{2}[r_x(t) -i
r_y(t)].
\end{align}

\begin{figure}[h]
\includegraphics[width=7.3cm, height=3.4cm]{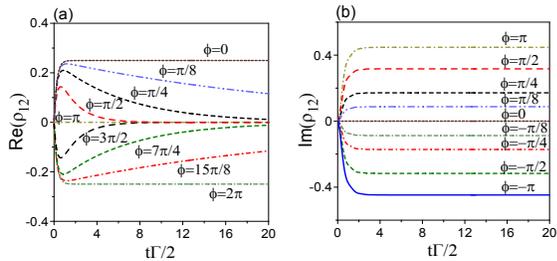}
\caption{The time evolution of $\rho_{12}$.  (a) ${\rm Re}\rho_{12}$
is plotted for $\phi$ from $0$ to $2\pi$ since it is antisymmetric
to $\phi=\pi$. (b) ${\rm Im}\rho_{12}$ is plotted for $\phi$ from
$-\pi$ to $\pi$ and it is antisymmetric to $\phi=0$. Here we take
$\mu_{L}-\mu_{R}=6\Gamma$ at temperature $k_{B}T=\Gamma/5$.
\label{fig2} }
\end{figure}

\begin{figure}[h]
\includegraphics[width=7.0cm, height=6.0cm]{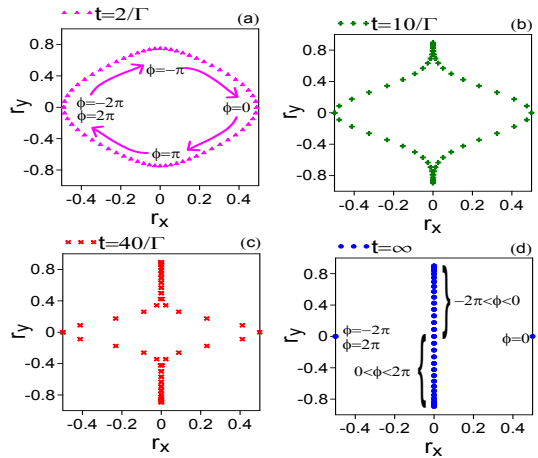}
\caption{Distribution of $(r_{x},r_{y})$ among different values of
$\phi$ from $-2\pi$ to $2\pi$ at different times. The parameters are
the same as that used in Fig.~\ref{fig2}. Note that changing the
direction of the magnetic flux $\phi\rightarrow-\phi$, one flips the
sign of $r_{y}$ while the sign of $r_{x}$ is flipped around
$\phi=\pm\pi$ as seen also in Fig.~\ref{fig2}. \label{fig3} }
\end{figure}

Fig.~\ref{fig2} plots the time evolutions $\rho_{12}$. At
$\phi=2m\pi$ where $m$ is an arbitrary integer, ${\rm Re}\rho_{12}$
soon grows to stable values $\pm1/4$ (with a time scale $\sim
1/\Gamma$) and ${\rm Im}\rho_{12}$ keeps zero, which locks the
relative phase $\varphi$ to $0$ or $\pi$. However, when
$\phi\ne2m\pi$, ${\rm Re}\rho_{12}$ also grows to a maximal value
(coherence building) with the same time scale and then decays to
zero at a flux dependent rate $\Gamma(1-|\cos(\phi/2)|)$. Eventually
all the different relative phases are localized to $\pi/2$ or
$-\pi/2$. Fig.~\ref{fig3} further visualizes this process.  At short
time $t=2/\Gamma$, the direction of ${\bm r}(t)$ sweeps over the
whole plane, which shows all kinds of relative phases between the
two charge states induced by different values of the magnetic fluxes
($-2\pi \le \phi \le 2\pi$) [see Fig.~\ref{fig3}(a)], as a process
of building coherence from the initial empty state.  These different
values (corresponding to different relative phases) then move toward
the $y$ axis as time goes on, except for the points $\phi=(0, \pm
2\pi)$, see Fig.~\ref{fig3}(b) and (c), and finally reside on it as
the asymptotic limit shown in Fig.~\ref{fig3}(d) as the phase
localization.

The period of the two charge states in $\phi$ is $4\pi$ as a result
of the intrinsic geometry of a two-level system.  Though the
relative phase is constant over the whole range of $-2\pi < \phi <
2\pi$, except for $\phi=0$, the degree of interference $|\rho_{12}|$
continuously changes with the flux as well as the bias.  At zero
bias, $|\rho_{12}|$ becomes zero in the steady limit, as a result of
the statistical equilibrium. Therefore, the nonzero bias is crucial
to manifest the above phase localization. We should also point out
that this phase localization over the magnetic flux is gauge
independent.

\begin{figure}[h]
\includegraphics[width=8.0cm, height=6.5cm]{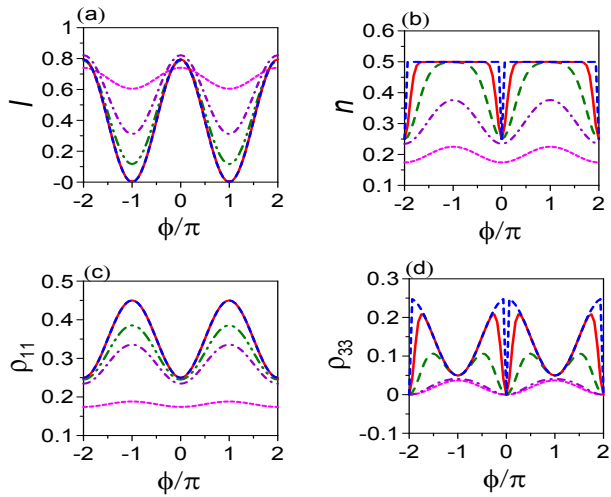}
\caption{The dependencies of the current $I$, the occupation number
$n=n_{1}=n_{2}=\rho_{11}+\rho_{33}$, single occupancy probability
$\rho_{11}$ and double occupancy $\rho_{33}$ on $\phi$ are examined
at various times (the line styles are $t=0.6/\Gamma$: pink
ultrashort dashed, $t=1.4/\Gamma$: purple dash-dot, $t=2/\Gamma$:
green dash-dot-dot, $t=6/\Gamma$: green long dashed, $t=40/\Gamma$:
red solid and $t=\infty$: blue short dashed). Both $\rho_{11}$ and
$\rho_{33}$ oscillate in flux in such a way that $n$ is kept as a
constant for $\phi\ne2m\pi$. \label{fig4} }
\end{figure}

To observe such process, we examine two measurable quantities, the
transport current passing through the double dot and the occupation
number in the two dots. The former, usually derived from the
nonequilibrium Green function technique \cite{Hau98}, can also be
obtained within our theory (see the explicit derivation in
\cite{Jin10083013} and the result is consistent with \cite{Hau98}):
\begin{align} I(t)=\frac{\Gamma}{2}{\rm
Re}\int \frac{d\omega}{2\pi}
(f_{L}(\omega)-f_{R}(\omega))\Big\{(1+|\cos\frac{\phi}{2}|)u_{+}(t,\omega)\notag
\\
+(1-|\cos\frac{\phi}{2}|)u_{-}(t,\omega)-\Gamma
\sin^2\frac{\phi}{2}u_{+}(t,\omega)u^{*}_{-}(t,\omega)\Big\}.
\end{align}
The last term in the above current expression is indeed proportional
to ${\rm Im}\rho_{12}$.  The result is plotted in
Fig.~\ref{fig4}(a). As one can see, the current always oscillates
smoothly in $\phi$ (showing the AB oscillation alone). The sharp
transitions across $\phi=2m\pi$ and the invariance of the relative
phase over the whole range of flux, as the main features of phase
localization, do not be seen in the transport current. However, we
find that the occupation number, $n_{i}=\rho_{ii}+\rho_{33}$ with
$i=1,2$, which depends on ${\rm Re}\rho_{12}$ via
$\dot{n}_{i}(t)=-\Gamma n_{i}(t)-\Gamma\cos(\phi/2){\rm
Re}\rho_{12}(t)-\tilde{\gamma}_{ii}(t)$, does display the features
of the phase localization discussed above. The occupation number
initially oscillates smoothly in $\phi$, and in the steady limit,
\begin{align}
n_{i}(t\rightarrow\infty)={1\over2}-\cos(\phi/2){\rm
Re}\rho_{12}(t\rightarrow\infty).
\end{align} As shown in Fig.~\ref{fig4}(b),
the phase localization to $\pi/2$ and $-\pi/2$ [as a result of ${\rm
Re}\rho_{12}(t\rightarrow\infty)= 0$] are explicitly manifested by
the invariance of $n_{i}(t\rightarrow\infty)=1/2$ over the whole
range of the flux, except for $\phi=2m\pi$ ($m=0, \pm1$). At
$\phi=2m\pi$, sharp change of ${\rm Re}\rho_{12}$ to $\pm1/4$ is
also manifested as a sharp reduction of the occupation number
$n_{i}$ from $1/2$ to $1/4$.

Fig.~\ref{fig4}(c) and (d) further reveals the underlying picture of
the phase localization. It shows that $\rho_{11}$ only contains the
AB oscillation (oscillates in $\phi$ smoothly), just like the
transport current $I$ does [see Fig.~\ref{fig4}(a) and (c)]. The
asymptotic constant occupation $1/2$ (manifesting the phase
localization) for any flux $\phi \neq 2m\pi$ is contributed mainly
from the double occupation $\rho_{33}$. Furthermore, the sharp
reduction of occupation number from $1/2$ to $1/4$ is indeed rooted
in the vanishing double occupancy at $\phi=2m\pi$. As we know,
strong inter-dot Coulomb repulsion may prohibit double occupancy
regardless of the flux. Phase localization then might not be
expected in this case. However, the parameters of the system can be
well tuned to make the inter-dot Coulomb repulsion become
unimportant so that the phase localization can always be manifested.

\begin{figure}[h]
\includegraphics[width=8.0cm, height=3.3cm]{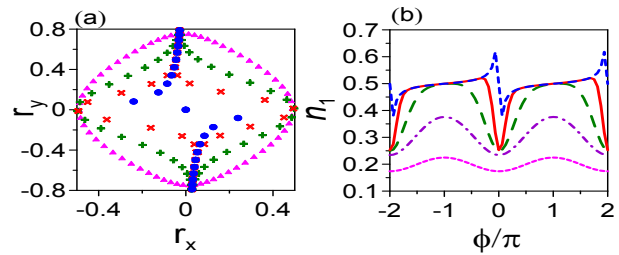}
\caption{The distribution of the polarization $(r_{x},r_{y})$ and
the occupation number $n_{1}$ deviated away from the degenerate
double dot, with energy level difference $E_{1}-E_{2}=2.5\%\Gamma$,
are plotted
 in (a) and (b), respectively, at various times. The symbols and the
 corresponding time values used in (a) are the same as that used in
 Fig.~\ref{fig3}.
The different curves corresponding to various times in (b) are the
same as that in Fig.~\ref{fig4}(b).  $n_{2}$ is almost the same as
$n_{1}$ since the deviation is small.   \label{fig5} }
\end{figure}

The above discussion has centered on degenerate double dot.
Practically there is no perfect degenerate double dot. However, we
find that slightly splitting the degeneracy only slightly deviates
the result of phase localization while the main property is still
well preserved.  As one can see from Fig.~\ref{fig5}(a),
$r_{x}=2{\rm Re}\rho_{12}$ is lifted a little bit from zero for the
flux values near $2m\pi$.  Similar signals on the occupation number
can also be seen reflecting such a slight deviation for the phase
localization, as shown in Fig.~\ref{fig5}(b). Comparing to the
transport current, the electron occupation in each dot is easier to
be measured in experiments using, for example, SET or QPC
\cite{Devoret}, this phenomenon can be well observed experimentally.

\emph{Conclusion.---} We have directly studied the dynamical effects
of the magnetic flux on the relative phase between two electron
charge states in a double dot AB interferometer by exactly solving
the nonequilibrium electron dynamics of the system. We found that
the flux dependence of the relative phase characterizing the
intrinsic electron coherence lead to a phase localization through
the nonequilibrium electron transport over the whole range of flux
except for the points $\phi=2m\pi$ ($m=0, \pm1)$. We also found that
such distinguished dynamics of phase localization is manifested in
the occupation number which is expected to be measurable in
experiments.

\begin{acknowledgments}
We would like to thank Amnon Aharony and Ora Entin-Wohlman for the
extensive and fruitful discussions to the problems addressed in this
paper and also for helping to improve the presentation of the
manuscript. This work is supported by the National Science Council
of ROC under Contract No. NSC-99-2112-M-006-008-MY3. We also thank
the support from National Center for Theoretical Science of Taiwan.
\end{acknowledgments}


\begin{thebibliography}{999}
\bibitem{COM1} It can be shown \cite{Jin10083013} that  ${\bm u} $ and ${\bm v}$ are
directly related to the retarded and lesser Green functions,
respectively,  of the nonequilibrium Keldysh scheme \cite{Hau98}.
\end{thebibliography}
\end{document}